\documentclass{aa}
\pdfoutput=1

\usepackage{natbib}
\usepackage{graphicx}
\usepackage{color}
\usepackage{txfonts}
\usepackage{hyperref}
\hypersetup{colorlinks, allcolors=blue}

\usepackage[T1]{fontenc}
\usepackage{verbatim}
\usepackage{textcomp}
\usepackage{textgreek}

\newcommand{\sig}{$\sigma$~}
\newcommand{\m}{$^\mathrm{m}$~}
\newcommand{\um}{$\mu$m~}
\newcommand{\ume}{$\mu$m}
\newcommand{\msun}{M$_\odot$~}
\newcommand{\msune}{M$_\odot$}
\newcommand{\s}{$\sim$}
\newcommand{\h}[1]{$^{#1}$}
\newcommand{\spa}{stars arcsec$^{-2}$~}
\newcommand{\spae}{stars arcsec$^{-2}$}

\usepackage{amstext}

\begin{document}

  \title{The Future of IMF studies with the ELT and MICADO\thanks{This work uses the instrument simulation package for MICADO, SimCADO: \url{https://simcado.readthedocs.io/}}}
  \subtitle{I: The local Universe as a resolved IMF laboratory}
  \author{K. Leschinski\inst{1}
     \and
          J. Alves\inst{1}
     }

  \institute{University of Vienna, Department of Astrophysics,
          Vienna, Austria\\
          \email{kieran.leschinski@univie.ac.at}
     }

  \date{Received 14.04.2020; accepted 13.05.2020}

%%%%%%%%%%%%%%%%%%%%%%%%%%%%%%%%%%%%%%%%%%%%%%%%%%%%%%%%%%%%%%%%%%%%%%%%%%%%%%%%
% ABSTRACT
%%%%%%%%%%%%%%%%%%%%%%%%%%%%%%%%%%%%%%%%%%%%%%%%%%%%%%%%%%%%%%%%%%%%%%%%%%%%%%%%

  \abstract
{Young stellar cluster cores in the local Universe provide the most pristine information available on the stellar initial mass function (IMF), but their stellar densities are too high to be resolved by present-day instrumentation.
With a resolving power 100 times better than the Hubble Space Telescope, the Multi-Adaptive Optics Imaging CameraA for Deep Observations (MICADO), which is the near-infrared camera on the Extremely Large Telescope (ELT), will for the first time provide access to a significant number of dense young stellar clusters that are critical for direct studies on the universality and shape of the IMF.}
{In this work we aim to estimate the lowest stellar mass that MICADO will be able to reliably detect given a stellar density and distance.
We also show that instrumental effects that will play a critical role, and report the number of young clusters that will be accessible for IMF studies in the local Universe with the ELT.}
{We used SimCADO$^*$, the instrument simulator package for the MICADO camera, to generate observations of 56 dense stellar regions with densities similar to the cores of young stellar clusters.
We placed the cluster fields at distances between 8\,kpc and 5\,Mpc from the Earth, implying core densities from 10\h2 to 10\h5\,\spae, and determined the lowest reliably observable mass for each stellar field through point-spread function (PSF) fitting photometry.}
{Our results show that stellar densities of \textless10\h3\,\spa will be easily resolvable by MICADO. The lowest reliably observable mass in the Large Magellanic Cloud will be around 0.1\,\msun for clusters with densities \textless10\h3\,\spae.
MICADO will be able to access the stellar content of the cores of all dense young stellar clusters in the Magellanic Clouds, allowing the peak and shape of the IMF to be studied in great detail outside the Milky Way.
At a distance of 2\,Mpc, all stars with M\,\textgreater\,2\,\msun will be resolved in fields of \textless10\h4\,\spa, allowing the high-mass end of the IMF to be studied in all galaxies out to and including NGC\,300.}
{We show that MICADO on the ELT will be able to probe the IMF of star clusters that are ten times denser than what the James Webb Space Telescope will be able to access,
% --- re-writen for clarity ---
and over one hundred times denser than the clusters that the Hubble Space Telescope can successfully resolve.
While the sensitivity of MICADO will not allow us to study the brown dwarf regime outside the Milky Way, it will enable access to all stellar members of over 1000 young clusters in the Milky Way and the Magellanic Clouds.
Furthermore, direct measurements of the Salpeter slope of the IMF will be possible in over 1500 young clusters out to a distance of 5\,Mpc.
MICADO on the ELT will be able to measure resolved IMFs for a large ensemble of young clusters under starkly different environments and test the universality of the IMF in the local Universe.}


\maketitle

%@arxiver{5_clusters.pdf,old_trusted_mass.pdf,resolved_stellar_densities.pdf}

%________________________________________________________________

%%%%%%%%%%%%%%%%%%%%%%%%%%%%%%%%%%%%%%%%%%%%%%%%%%%%%%%%%%%%%%%%%%%%%%%%%%%%%%%%
% INTRODUCTION
%%%%%%%%%%%%%%%%%%%%%%%%%%%%%%%%%%%%%%%%%%%%%%%%%%%%%%%%%%%%%%%%%%%%%%%%%%%%%%%%

\section{Introduction}
\label{sec:introduction}

\begin{table*}

    \centering
    \caption{Mass limits for a selection of studies of the IMF outside the Milky Way with the Hubble Space Telescope. }
    \label{tbl:imf_lit_review}

    \begin{tabular}{ l l r r r r c }
        \hline
        \hline
        Galaxy   &  Target      &  Distance &  Mass range       & IMF slope(s) & Break mass          & Reference         \\
                &               & kpc       & \msun             &              & \msun               &                   \\
        \hline
        LMC      &  R136        & 50        & 2.8--15            & 2.22         &                     & 1 \\
        LMC      &  NGC 1818    & 50        & 0.85--9            & 2.23         &                     & 2  \\
        LMC      &  R136        & 50        & 1.35--6.5          & 2.28, 1.27   & 2.1                 & 3     \\
        LMC      &  LH 95       & 50        & 0.43--20           & 2.05, 1.05   & 1.1                 & 4      \\
        \hline
        SMC      &  NGC 330     & 62        & 1--7               & 2.3          &                     & 5    \\
        SMC      &  NGC 602     & 62        & 1--45              & 2.2          &                     & 6     \\
        SMC      &              & 62        & 0.37--0.93         & 1.9          &                     & 7     \\
        \hline
        Hercules &              & 135       & 0.52--0.78         & 1.2          &                     & 8         \\
        Leo IV   &              & 156       & 0.54--0.77         & 1.3          &                     & 8         \\
        Leo I*   &              & 250       & 0.6--30            & 2.3          &                     & 9     \\
        \hline
        \end{tabular}

        \tablefoot{For the study by \citet{gallart1999}, the estimated global star formation history was consistent with a Salpeter slope. The Salpeter slope was not extracted from the photometric data.}
        \tablebib{(1)~\citet{hunter1995}; (2) \citet{hunter1997}; (3) \citet{sirianni2000}; (4) \citet{dario2009};
                  (5) \citet{sirianni2002}; (6) \citet{schmalzl2008}; (7) \citet{kalirai2013}; (8) \citet{geha2013}
                  (9) \citet{gallart1999}.
                }

\end{table*}

The stellar initial mass function (IMF), or the spectrum of stellar masses at birth, has implications in almost all fields of astrophysics.
%On the local scale, the IMF determines the number of available massive stars and with it the fate of a star formation region and creating\LEt{it is not clear what "and with it" and also "and creating" belongs to. Do you mean "the IMF determines the number of available massive stars. This number holds information about the fate of a star formation region. Massive stars create the environments..."? Please check and change for clarity} the environments that emergent planet-forming circumstellar discs will be exposed too.
On the local scale, the IMF describes the number of massive stars that will form during a star-forming event.
This dictates both the ultimate fate of a star formation region and the environment that emergent planet-forming circumstellar discs will be exposed too.
On the large scale, the IMF is irrevocably connected to the composition of the stellar populations in a galaxy and has a critical effect on the mass and energy cycle of a galaxy.
The larger the amount of mass locked up in low-mass stars, the smaller the reservoir of gas available for the next generation of stars, and consequently, the lower the potential for the enrichment of the interstellar medium (ISM).
Finally, cosmological simulations of galaxy formation and large-scale structure inevitably rely on a universal IMF to determine stellar yields and the strength of feedback mechanisms governing the transport of energy and material.

In his original work, \citet{salpeter1955} used a single power-law distribution with a slope of 2.35 to describe the IMF for masses between \s1\ and \s10\,\msune.
This description was later modified to a series of broken power laws to include the stars below the hydrogen-burning limit by \citet{kroupa2001}\@.
\citet{chabrier2003, Chabrier2005} proposed a log-normal distribution with a power-law modification for the high-mass regions.
Given the observational uncertainties involved in determining the exact shape of a cluster's IMF, it has hitherto proved difficult to show which of these two descriptions more aptly describes the IMF\@.

Most observational studies suggest that the shape of the IMF is constant~\citep{Lada2003-ip,Kroupa2002,Bastian2010}.
Definitive deviations from the accepted IMF form are elusive, and when found, are often controversial~\citep{Van_Dokkum2010-gx,Conroy2012-hv,Drass2016-kp}.
% Rewritten for clarity
However what currently hinders a conclusive result regarding the universality of the IMF is the lack of a statistically relevant sample of directly derived IMFs in environments substantially different from the solar neighbourhood.
% One main challenge\LEt{this is not quite clear. The challenge is not the directly derived IMFs, but that it is difficult to obtain them, correct? Please rephrase this to reflect that, maybe like "We cannot conclude on the universality of the IMF without directly derived IMFs..." or similar} hinders a concluding result on the universality of the IMF:  directly derived IMFs from star-counts for environments that are substantially different from the solar neighbourhood.
Table~\ref{tbl:imf_lit_review} shows that even in the closest star-forming galaxies like the Magellanic Clouds, only the Hubble Space Telescope (HST) has the sensitivity to reach below one solar mass (see references in Table~\ref{tbl:imf_lit_review}).
Long exposures with the HST have observed stars just below the first break in the Kroupa power law at 0.5\,\msun~\citep{dario2009,kalirai2013,geha2013}, but not far enough into the lower mass regions to place reliable constraints on the shape of the IMF in these extragalactic environments.
Adding to observers' woes is the lack of spatial resolution.
At the distance of the Large Magellanic Cloud (LMC), star-forming regions can contain anywhere from 10\h2 to 10\h5\,\spae.
Figure~1 of \citet{sirianni2000} shows a perfect example of why current studies struggle to reliably determine the IMF for dense stellar populations outside the Milky Way.
The depicted cluster core (R136) is completely dominated by the flux of a few of the brightest stars.
Studies of the IMF are thus limited to the outer regions of these clusters where stellar densities are low enough for individual low-mass stars to be resolved.
Unconstrained mass segregation can also skew the results when investigating the IMF of massive clusters (e.g., \citealt{Ascenso2009-de}).
More importantly, because current telescopes are unable to resolve the IMFs of massive clusters both in- and outside the Milky Way, strong assertions on the universality of the IMF are problematic.
In order to systematically and unambiguously study (through star counts) the lower mass part of the IMF and to be able to characterise differences between IMFs, telescopes with higher spatial resolution and better sensitivity than the current generation of ground- and space-based telescope are needed.

In the middle of the next decade, the era of the extremely large telescopes will begin.
ESO's Extremely Large Telescope (ELT)~\citep{eelt} will with the help of advanced adaptive optics (AO)~\citep{maory} have the power to resolve spatial scales at the diffraction limit of a 40m class mirror.
This will provide a linear improvement of a factor of \s15$\times$ over HST and a factor of \s6$\times$ over the future James Webb Space Telescope (JWST).
With a collecting area of 978\,m\h2, the ELT will have at least the same sensitivity as the HST in a sparse field, and it will be able to observe much deeper than HST in crowded fields.
The Multi-Adaptive Optics Imaging CameraA for Deep Observations (MICADO) ~\citep{micado2016, micado2018} will be the first-light near-infrared (NIR) wide-field imager and long-slit spectrograph at the ELT\@.
With a diffraction limit of 7\,mas at 1.2\um and an AO-corrected field of view of almost a square arcminute, MICADO will be perfectly suited to address the IMF science case.

The main focus of this paper is to determine to which extent MICADO will improve our ability to study the IMF and other properties of dense stellar populations.
% Rewritten to remove the numbered list
More precisely, we determine the lowest mass star that MICADO will be able to observe for a given density and distance.
We also shed light on the instrumental effects that will play a critical role when such studies are undertaken with MICADO and the ELT.
As theoretical work is always best connected to an observable quantity, we also present an estimate of the number of young clusters that will be available for IMF studies with the ELT\@ out to a distance of 5\,Mpc.
%More precisely, we address the following three questions\LEt{direct questions in a paper are unusual. Please rephrase along the lines of "... we determine the lowest mass of a star that MICADO will be able to observe ..." The questions look too much like the work order that prompted the paper, and the paper itself should be better than an order list}:
%1) What is the lowest mass star that MICADO will be able to observe for a given density and distance?
%2) What instrumental effects will play a critical role when such studies are undertaken with MICADO and the ELT? and
%3) How many young clusters will be available for IMF studies with the ELT?

In our quest for answers, we used SimCADO, the instrument data simulator for MICADO~\citep{leschinski2016, leschinski19}, to simulate a wide range of densely populated stellar fields at various distances.
The current version of SimCADO takes all the major and most of the minor spatial and spectral effects along the line of sight between the source and the detector into account.
We used the software to generate realistic images of model stellar fields and  conducted several iterations of point-spread function (PSF) photometry and star subtraction to extract as many stars as possible from the simulated observations.
The extracted stars were compared with the input catalogue to determine the completeness of the extraction and to define a limiting reliably observable mass for the different stellar field densities and distances.

This paper follows several recent works on resolved stellar populations with extremely large telescopes
(e.g., \citealt{deep11, greggio12, gullieuszik15, tolstoy19_iau}).
A brief overview of the main MICADO science cases is given in~\cite{micado2016}.
The latest science cases and simulations for the InfraRed Imaging Spectrograph (IRIS) on the Thirty Meter Telescope (TMT) can be found in~\cite{tmt_iris16}.
% The latest science cases and simulations for\LEt{please provide the spelled-out versions of IRIS and (TMT) and all other abbreviations at first occurrence, as I did for you so far. I am not going to continue this to get on with the editing proper} IRIS on the TMT can be found in~\cite{tmt_iris16}.

This paper is organised in the following way: Section~\ref{sec:observations} describes the stellar fields we used in our simulations, the simulations themselves, and the algorithm for detecting and subtracting stars in the simulated images.
In Section~\ref{sec:results} we describe the results of the simulations and discuss their validity in the context of possible future observations of real young stellar clusters.
Section~\ref{sec:conclusion} summarises our results.

%%%%%%%%%%%%%%%%%%%%%%%%%%%%%%%%%%%%%%%%%%%%%%%%%%%%%%%%%%%%%%%%%%%%%%%%%%%%%%%%
% OBSERVATIONS / DATA SET
%%%%%%%%%%%%%%%%%%%%%%%%%%%%%%%%%%%%%%%%%%%%%%%%%%%%%%%%%%%%%%%%%%%%%%%%%%%%%%%%

\section{Data sets}
\label{sec:observations}

\begin{figure*}
    \centering
    \includegraphics[width=\textwidth]{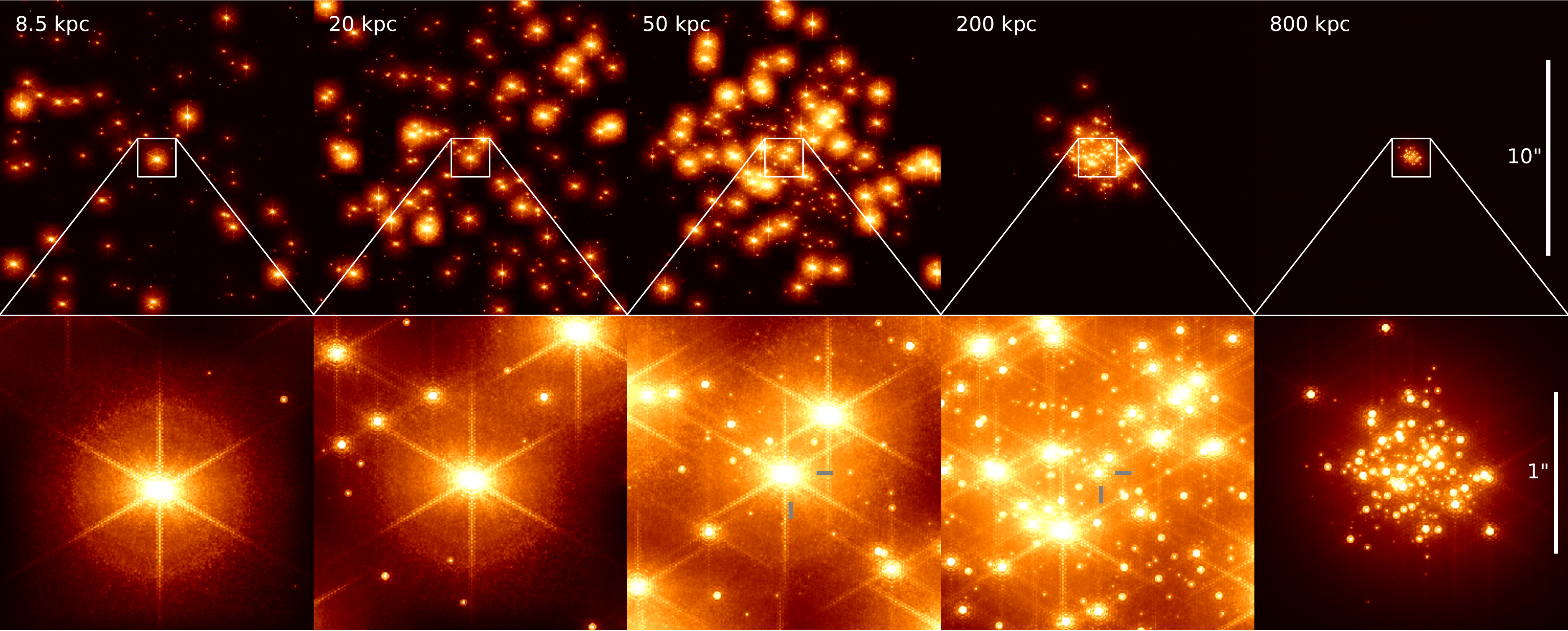}
    \caption{Illustration of the synthetic data.
    The model young massive cluster depicted here contains a mass of 2$\times$10$^4$\,\msun ($\sim$5$\times$10$^4$ stars) and has a half-light radius of 1\,pc.
    It contains stars in the mass range [0.01, 100]\msune, sampled from a Kroupa IMF.
    The model cluster was observed with the MICADO instrument simulator at distances from 8\,kpc to 800\,kpc.
    The top row shows the cluster as seen by the central detector of MICADO (covering an on-sky area of 16\arcsec$\times$16\arcsec).
    The bottom row shows a 2\arcsec$\times$2\arcsec window (512$\times$512 pixels) at the centre of this detector.
    This is the area we used for our study.
    Here the unique structure of the ELT PSF is visible.
    This figure illustrates the difficulties that observers will face when crowded fields are studied with MICADO and the ELT.
    }
    \label{fig:5_clusters}
\end{figure*}

\begin{figure*}
    \centering
    \includegraphics[width=\textwidth]{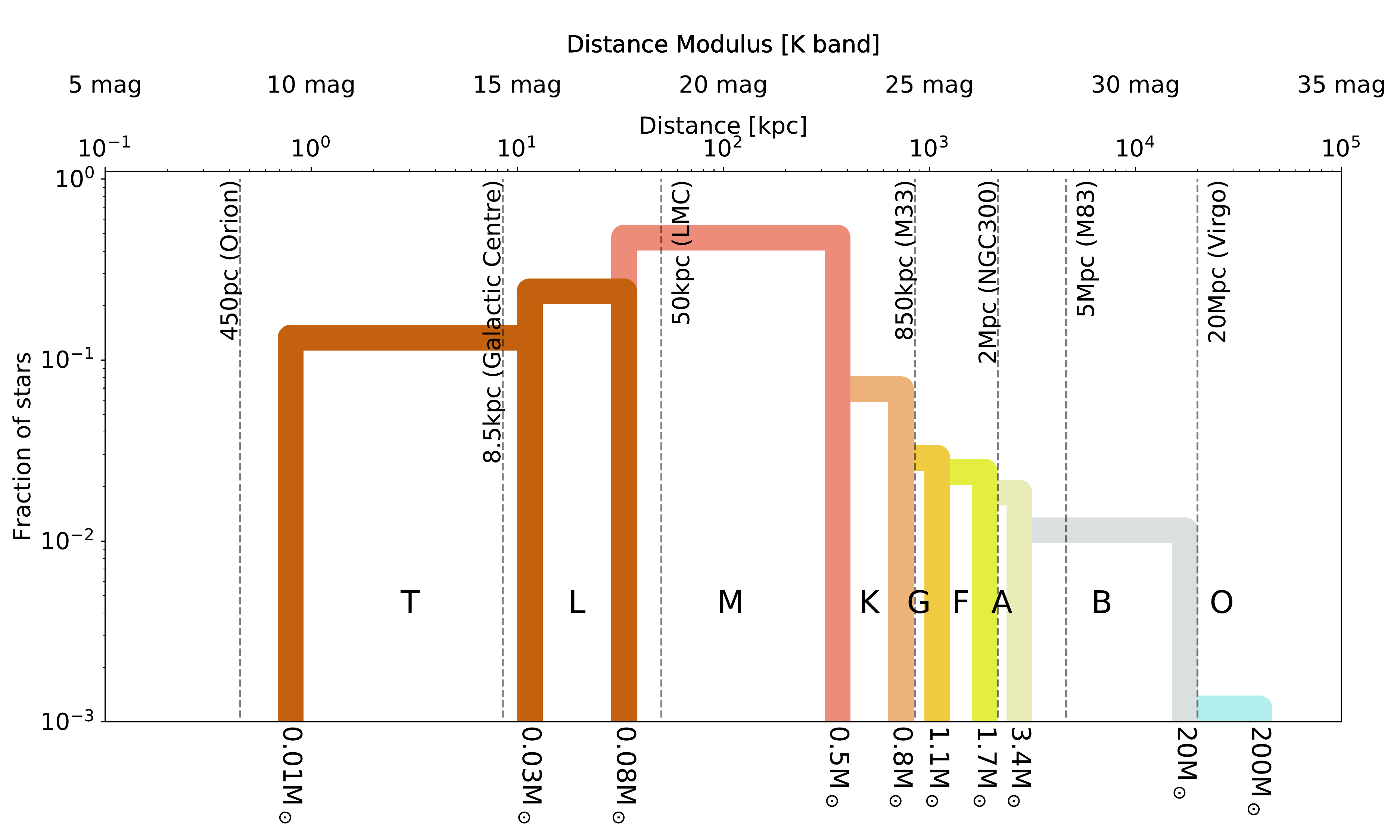}
    \caption{Qualitative illustration of the observational horizons of main-sequence spectral types assuming a sensitivity limit of K$_S$=28$^m$ with MICADO and the ELT.
    The vertical dotted lines show the distance moduli of regions that are important for studying the IMF in environments that are significantly different from the solar neighbourhood.
    The height of each bar corresponds to the cumulative fraction of all stars belonging to a specific spectral type.
    }
    \label{fig:imf_educational}
\end{figure*}

\begin{figure*}

    \centering
    \includegraphics[width=\textwidth]{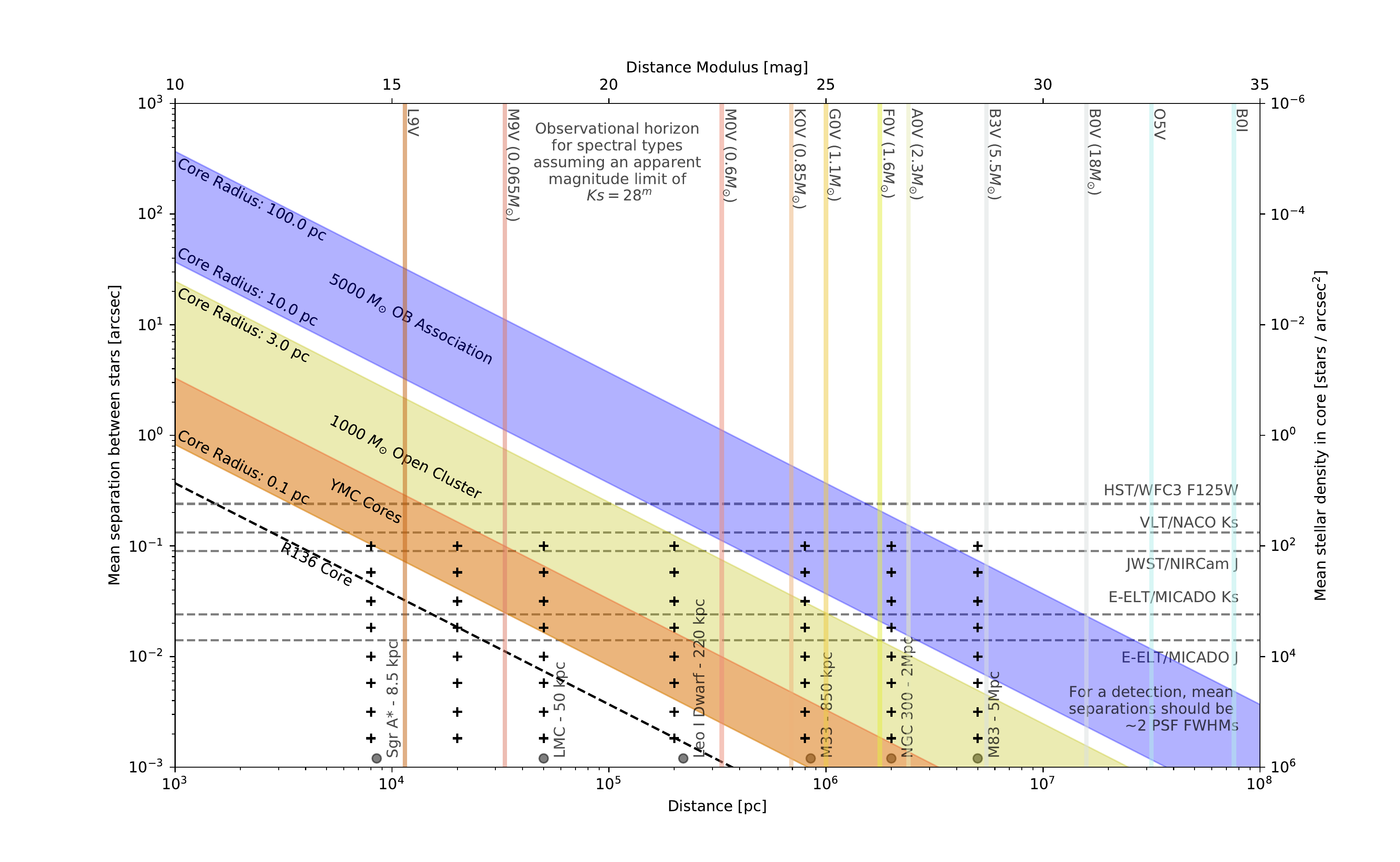}

    \caption{Stellar density parameter space covered by the dense stellar fields in this study (crosses).
    The diagonal bands represent the range of core densities for the three main categories of young stellar populations: YMCs (orange), open clusters (green), and OB associations (blue).
    The vertical lines represent the farthest distance at which a particular type of main-sequence star will still be above the detection limit of MICADO, i.e., Ks=28\m.
    The dashed horizontal lines show the theoretical confusion limit for MICADO/ELT, JWST, HST, and an instrument similar to NACO/VLT.
    The confusion limit assumes an average minimum distance of twice the PSF FWHM between stars.
    }

    \label{fig:resolved_stellar_densities}

\end{figure*}

The most reliable way to determine the IMF is to study a population of stars that is still young enough for all (or most) of its original members to still be around, but that is old enough for the main phase of star formation activity to have ceased.
If a population is too young, it will not have completed forming all its stars, and dust extinction will be a significant source of uncertainty and incompleteness.
  If it is too old, the most massive members will already have exploded as supernovae.
Dynamical effects will also have led to the evaporation of stars from the cluster, posing a problem to IMF completeness.
Unfortunately, such ideal conditions are rarely found.
Star formation occurs on timescales of 10\h6 years.
The most massive stars burn their hydrogen reserves within the first few to several ten million years and move off the main sequence.
Because the dispersion time for stellar clusters is on the order of hundreds of millions of years (e.g., \citealt{Lada2003-ip}), at any point in time, relatively few of the observable new clusters will be found in the ideal age range (between 5 and 20\,Myr) for studying the IMF\@.
The majority of IMF studies focus on clusters that come closest to meeting these conditions: the cores of open clusters and young massive clusters (YMC).
OB associations also provide a laboratory for studying the IMF, with the advantage that stellar density does not pose a problem.
However, as OB associations are spread out over distances as vast as a few hundred parsecs, the probability of contamination from background sources is higher.

\subsection{Parameter space}
\label{subsec:parameter_space}

The HST has a diffraction limit of \s0.1\arcsec at 1.2\,\um and can reach magnitudes as faint as J=28.6\m~\citep{hst_wfc3} in a 10-hour observation.
With AO-assisted ground-based instruments on 8\,m class telescopes (similar to NACO at the VLT), diffraction-limited observations can be achieved over small (\s1\arcmin) fields of view.
The diffraction limit of the VLT (\s0.03\arcsec at 1.2\,\um) is about three times smaller than that of the HST, but because of the atmospheric background, the sensitivity limits of VLT instruments are many magnitudes brighter than those of the HST\@.
As boundary conditions for our suite of stellar fields, we took the resolution limit of the HST because cluster cores with densities lower than this are already accessible to the HST\@.
Assuming an average of one star per full width at half-maximum (FWHM) of the PSF, our lower density limit was set to 100\,\spae.
For the upper density limit, we first took the theoretical diffraction limit of the ELT: 7\,mas at 1.2\,\ume, or 2$\times$10\h4\,\spae.
Because we were looking to investigate crowding-limited observations at large distances (\textgreater1\,Mpc), where the faint stars had already dropped below the sensitivity limit, we increased the true stellar density by a factor of 15$\times$.
This meant that the crowding criterion of one star per FWHM was met even for simulated fields where only stars with masses M\,\textgreater\,1\,\msun were above the MICADO sensitivity limit.
We therefore set the upper limit for the true cluster stellar density to 3$\times$10\h5\,\spae.

Current telescopes are capable of detecting almost all main-sequence stars above the hydrogen-burning limit (\s0.08\,\msune) within a few kiloparsec of the Sun (e.g., \citealt{muzic17}).
Detecting all main-sequence stars in clusters farther away, for instance, in the galactic centre and beyond, is where the increased sensitivity and resolution of MICADO will bring the most significant breakthroughs.
The question of whether the IMF is truly universal indeed dictates that we study the IMF outside the Milky Way.
Figure~\ref{fig:imf_educational} illustrates the regions of the IMF that will be available to MICADO at various distances from Earth assuming a sensitivity limit of K$_S$=28$^m$.
We therefore placed our model proxy-clusters at distances corresponding to important regions that will be critical for studying the effect of the interstellar environment on the IMF: the Galactic centre (\s8\,kpc), the LMC (\s50\,kpc), the Leo I dwarf galaxy (\s200\,kpc), M33 (\s850\,kpc)\footnote{We recognise that the location of the ELT in the Southern Hemisphere is unfavorable for effectively observing M33.
We provide this data point because M33 will be observable by the Thirty Meter Telescope.}, NGC 300 (\s2\,Mpc), and M83 (\s5\,Mpc).
Figure~\ref{fig:resolved_stellar_densities} shows the parameter space covered by open clusters with an average mass (\s1000\,\msun) with radii between 0.1\,pc and 3\,pc and OB associations with an average mass (\s5000\,\msun) with radii between 10\,pc and 100\,pc with increasing distance from Earth.
The lower bounds of the open cluster parameter space also cover the cores of YMCs. Average cluster properties were derived for the OB associations from \citet{melnik1995}, for the open clusters from \citet{piskunov2007}, and for the YMCs from \citet{portegies2010}.

\subsection{Artificial stellar fields}
  \label{subsec:stellar_fields}
Figure~\ref{fig:5_clusters} shows a model cluster placed at ever-increasing distances from Earth.
It is evident from Figure~\ref{fig:5_clusters} that placing a single cluster at different distances would result in inhomogeneous data sets after running our source-finding algorithm.
In order to create a homogenous data set that would allow for a direct comparison of the effects of distance and the ELT optics on crowded fields, we generated 56 densely populated stellar fields that could function as proxies for the dense regions at the cores of young stellar clusters.
Each stellar field was generated for a unique combination of stellar density and distance.
The parameter space covered by these cluster proxies is shown by the crosses in Figure~\ref{fig:resolved_stellar_densities}.
The size of each stellar field was set at 2\arcsec$\times$2\arcsec (see section~\ref{sec:telescope}).
The stellar fields were populated by continually drawing stars from an IMF until the required stellar density was reached.
The mass of each star was drawn at random from an IMF distribution with minimum and maximum masses of 0.01\,\msun and 300\,\msun.
The IMF followed a standard \citet{kroupa2001} broken power-law distribution
\footnote{By ``standard'' we mean $\alpha=0.3$ for $\mathrm{M} < 0.08 \mathrm{M}_\odot$, $\alpha=1.3$ for $0.08\mathrm{M}_\odot < \mathrm{M} < 0.5 \mathrm{M}_\odot$, and $\alpha=2.3$ for $\mathrm{M} > 0.5 \mathrm{M}_\odot$ , as defined in Kroupa (2001)}.
Table~5 from \citet{pecaut2013}\footnote{Masses are not given in Table 5, but rather in the online supplement at \url{http://www.pas.rochester.edu/~emamajek/EEM_dwarf_UBVIJHK_colors_Teff.txt}}
was used to obtain the absolute J and Ks magnitudes for each star for the given mass.
The requisite distance modulus for the stellar fields was added to give each star an apparent magnitude.
We did not include extinction in the distance modulus because this varies with the line of sight, in particular for Milky Way clusters.
Because we study the worst-case scenario, the core of the massive clusters, the stars were assigned random coordinates within the 2\arcsec$\times$2\arcsec bounding box.
In contrast, real clusters will have a decreasing radial density profile outside the inner cluster core radius.
Characterising the IMF of clusters from real observations will therefore be more reliable than for our simulated images.

\subsection{Observations}
\label{sec:telescope}

To ``observe'' our synthetic stellar fields, we used the standard imaging mode of SimCADO\footnote{For documentation, see: \url{https://simcado.readthedocs.io/}.
Github code base: \url{https://github.com/astronomyk/SimCADO}}~\citep{leschinski2016}, an open-source instrument simulator that mimics observations with the wide-field mode of MICADO at the ELT\@.
The core regions of open clusters and YMC have radii on the order of 1\,pc~\citep{portegies2010}.
At a distance of 200\,kpc ($\sim$Leo 1 dwarf), this translates into an angular diameter of \s2\arcsec.
Thus we thought it safe to assume that the stellar density within the inner 2\arcsec$\times$2\arcsec region remains relatively constant.
To minimise computational effort, we decided to restrict to observations to this 2\arcsec$\times$2\arcsec window in the centre of the detector.

At the very least, dual-band photometry is required to determine the mass of a star.
Therefore detections in at least the J and K$_S$ filters are necessary.
We deem a detection in the K$_S$ filter to be critical for any study of the IMF and therefore restricted our observations to this filter.
The reason for this is as follows: The sky background in the K$_S$ filter is the highest of all near-infrared (NIR) filters, and the NIR stellar flux for all main-sequence stars (and many brown dwarfs) is weakest in the K$_S$ filter.
If a source is undetectable in the K$_S$ filter, it will not be possible to determine its mass accurately.
Based on the AO nature of the observations and the expected low Strehl ratio at 1.2\um~\citep{clenet2016}, it might be argued that detections in the J filter will be more difficult.
Ultimately, the fluxes of the stars and the sky are set by nature, whereas the Strehl ratio is a question of engineering and optical design.
The stars cannot be made brighter, whereas optical quality can be improved.
We therefore deemed a detection in the K$_S$ filter to be the critical point for determining the mass of cluster members.

Exposure times were kept to one hour for no other reason than observing time at the ELT will be in very high demand when it comes online, and observations in two or more filters are needed to accurately determine the mass of stars.

\subsection{Source extraction and matching}
\label{subsec:source_extraction}

Figures~\ref{fig:results_lmc_1E3} and~\ref{fig:results_lmc_1E4} in the appendix show a graphical representation of the process described in this section.
They show two examples of ``observed'' stellar fields placed at a distance of 50\,kpc and containing 10$^3$ and 10$^4$ \spa , respectively.
The stark features of the SCAO PSF are clearly visible in the images.
The diffraction core of the PSF, however, is still well modelled by a Gaussian distribution.
To find and measure the stars in the images, we used the following method:

\begin{enumerate}
    \item Find the brightest star in the image with \verb+DAOStarFinder+ from \verb+photutils+~\citep{photutils}
    \item Find the centre of the star in a 5$\times$5 pixel window around the coordinates given by \verb+DAOStarFinder+
    \item Fit a 2D Gaussian profile to the core of the star
    \item Scale an image of a reference star to match the amplitude, baseline, and offset of the detected star
    \item Subtract the scaled reference star from the image
    \item Repeat until \verb+DAOStarFinder+ no longer finds any sources above 5\,\sig
\end{enumerate}

In practice, we found that we were able to subtract about 100 stars at once and thus greatly increased the speed of the process.
The amplitudes and baselines were converted into magnitudes based on the reference star.
Our reference star was a solitary ``field'' star with a magnitude of K$_S$=15, observed for the minimum {MICADO} exposure time of 2.6\,s so as to maximise the signal-to-noise ratio while not saturating the detector.
We calculated the mass of each star based on the observed fluxes in the K$_S$ filter.
This step is only permissible because of the simplified context of this study.
We were free to equate the luminosity function with an equivalent mass function because all our stellar fields have the intrinsic property that they only contain main-sequence stars and the luminosity and mass functions enjoy a one-to-one relationship for this conversion in the K$_S$ filter, mathematically speaking.
Furthermore, our primary goal is to determine the lowest reliably observable mass, based on how well MICADO will perform in crowded fields. We do not intend to directly measure the mass of the original stars.
We are confident that this step does not significantly detract from achieving the goal of this study.

Finally, we cross-matched the coordinates of the extracted sources with the original table of coordinates to determine the fraction of stars that was correctly detected with our algorithm.
Because of noise and confusion from stars closer than a few FWHMs, the centroid coordinates of the extracted star were not always exactly equal to the original coordinates.
The cross-matching algorithm was instructed to search for the closest star within a 25\,mas radius.
If a fainter or brighter star happened to be closer, then the algorithm chose that star from the catalogue as the match.
We determined whether the extracted masses for stars in a certain mass bin were reliable by binning the extracted stars according to mass.
We then took the mean and standard deviation of all stars within a mass bin.
As long as the ratio of mean extracted mass to true mass was in the range 1.0$\pm$0.1 and the standard deviation was lower than 0.1, the mass bin was classed as reliable.
By this definition, the lowest reliably detectable mass for a stellar field was given by the lower edge of the lowest mass bin that satisfied these criteria.

%%%%%%%%%%%%%%%%%%%%%%%%%%%%%%%%%%%%%%%%%%%%%%%%%%%%%%%%%%%%%%%%%%%%%%%%%%%%%%%%
% RESULTS
%%%%%%%%%%%%%%%%%%%%%%%%%%%%%%%%%%%%%%%%%%%%%%%%%%%%%%%%%%%%%%%%%%%%%%%%%%%%%%%%

\section{Results and discussion}
\label{sec:results}

\subsection{Lowest reliably observable masses for given stellar densities and distances.}
  \label{subsec:lowest_mass}

\begin{figure*}

    \centering
    \includegraphics[width=\textwidth]{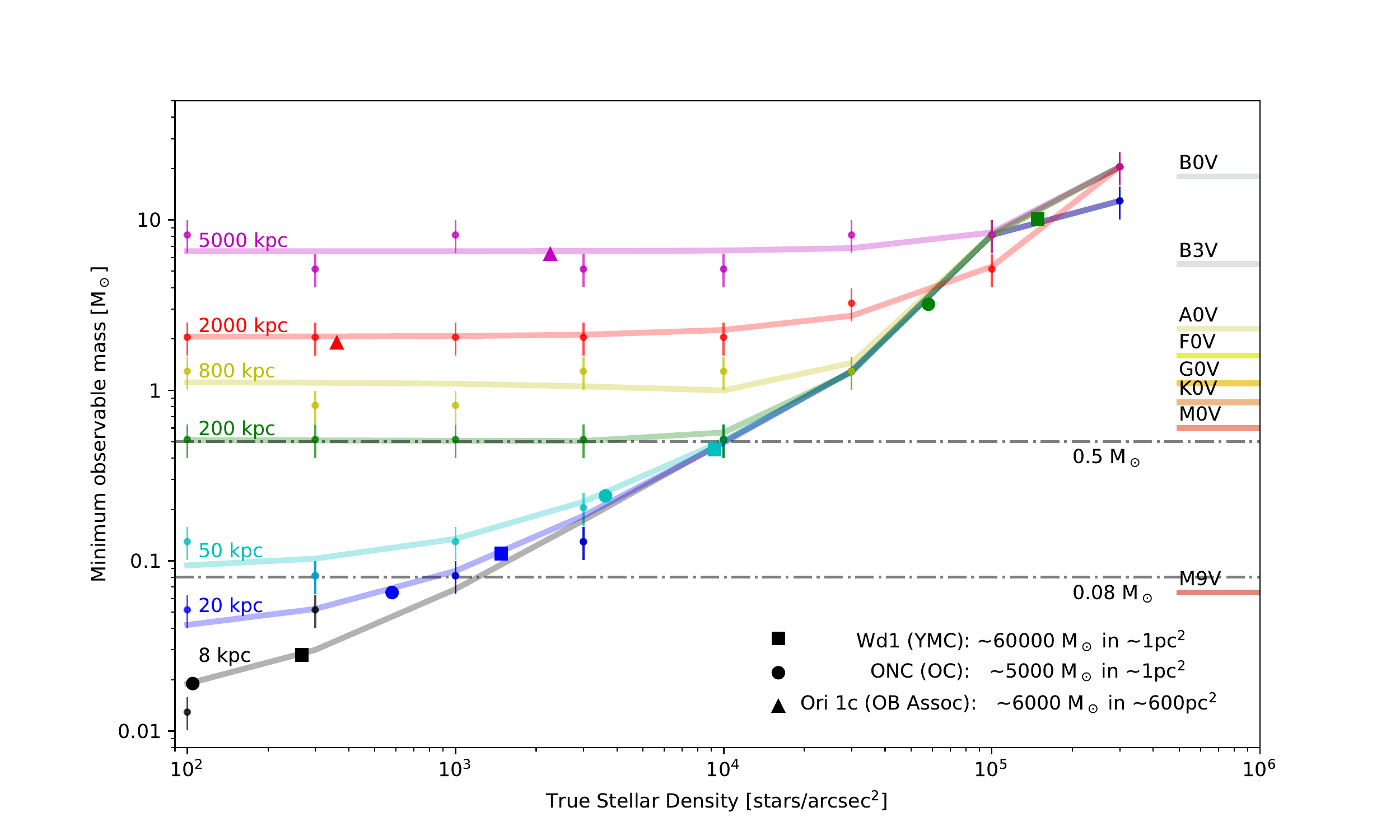}

    \caption{Lowest observable masses for given stellar densities and distances.
    The errors in the observable mass are 0.2 dex and correspond to the size of the mass bins.
    Two trends are visible in the best-fit lines for each distance: a flat regime where the limiting mass is constrained by the sensitivity limit of MICADO, and an exponential regime where crowding becomes the limiting factor.
    The void to the lower right shows the parameter space in which stars of a given mass will not be observable for a given stellar density.
    }
    \label{fig:trusted_mass}
\end{figure*}

\begin{figure*}

    \centering
    \includegraphics[width=\textwidth]{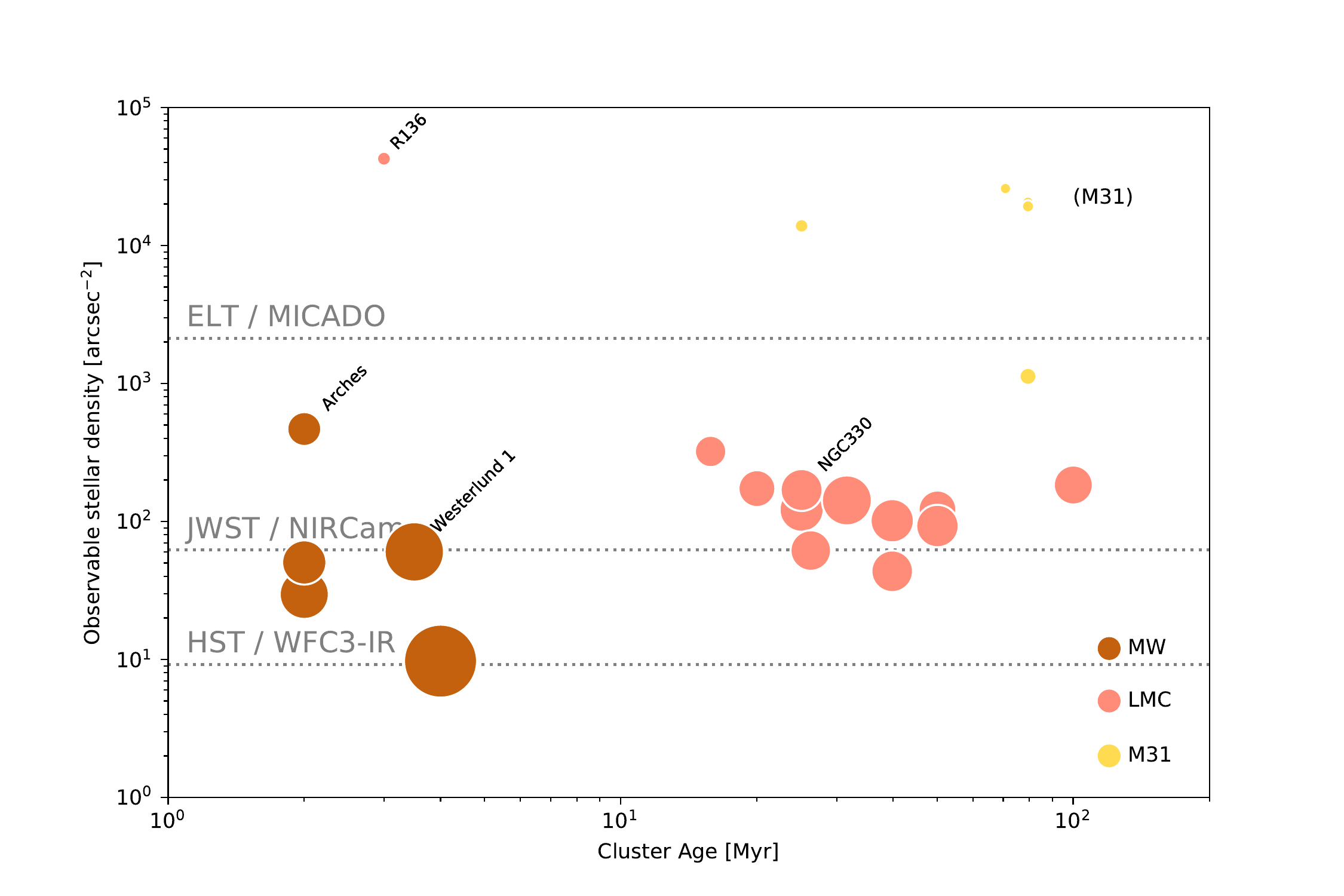}

    \caption{
    Stellar densities in the cores of the YMCs listed in Table~\ref{tbl:pz10_selection}, assuming a sensitivity limit of K$_S$=28\m.
    While these clusters are spread across the whole sky, only the handful of YMCs in M31 are outside the observing window of the ELT.
    The sizes of the circles roughly illustrate the relative on-sky sizes of the cluster cores.
    The colours reflect the minimum observable mass, as shown in Figure~\ref{fig:trusted_mass} and listed in Table~\ref{tbl:pz10_selection}.
    Brown represents M\textgreater0.01\msun, pink is for M\textgreater0.1\msun, and yellow shows M\textgreater0.9\msune.
    The densities shown here take the sensitivity limit into account and are therefore only for observable stars in each cluster, i.e, stars with K$_S$\textgreater28\m are omitted from the density calculation.
    This means that all stars in the Milky Way, M-type stars and brighter in the LMC, and G-type stars and brighter in M31, are included.
    The dashed lines in this figure represent the (estimated) limit to the resolving capability of the HST, JWST, and ELT.
    We define the limiting density as the density where the mean distance between stars is equal to twice the H-band PSF FWHM.
    Cluster cores of the majority of young clusters outside the Milky Way are far too dense for either HST or JWST observations. }

    \label{fig:star_density_vs_age}

\end{figure*}

\begin{figure*}

    \centering
    \includegraphics[width=\textwidth]{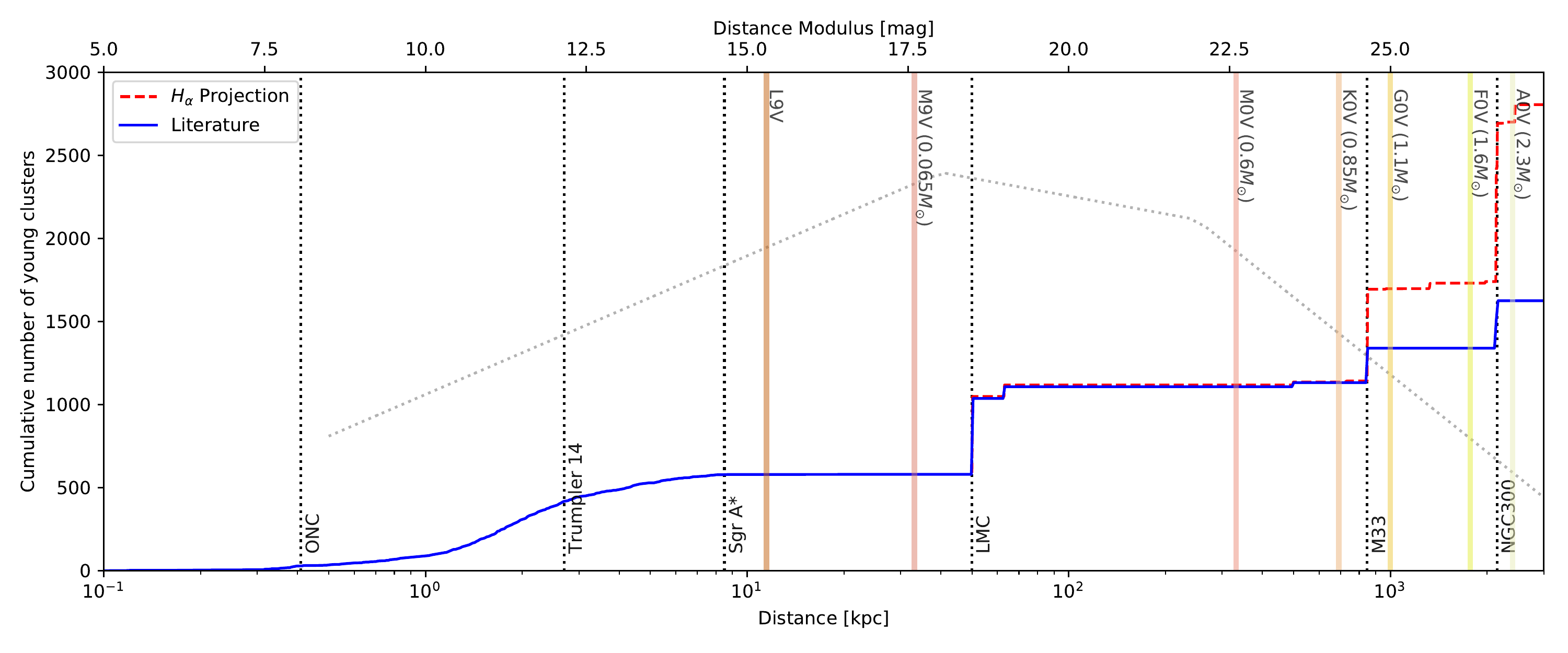}

    \caption{Cumulative number of young cluster targets that will be available to MICADO (\textdelta\,\textless\,+35\textdegree) out to 2 Mpc.
    The blue line shows the cumulative number of young clusters with increasing distance from Earth as reported in catalogues and the literature (see references in Table~\ref{tbl:cum_cluster_refs}).
    The red dotted line shows the expected number of young clusters in a given galaxy based on an extrapolation of a galaxy's star formation rate from its total H$_\alpha$ flux~\citep{caldwell09}.
    The solid vertical lines represent the observational horizons for given stellar spectral types assuming a detection limit of K=28$^m$ with MICADO at the ELT.
    The faint dotted grey line represents the regions of the IMF that are available to study with MICADO at different distances.
    The population of Milky Way clusters is taken from the HEASARC Milky Way Open Cluster database~\citep{heasarc_mwsc} and represents only the clusters located at declinations accessible to the ELT (-85\textdegree\,\textless\,\textdelta\,\textless\,+35\textdegree)
    }
    \label{fig:local_group_cluster_number}

\end{figure*}

The lowest mass star that MICADO will be able to observe reliably for a given density and distance constitutes the primary result of this study.
These masses are shown in Figure~\ref{fig:trusted_mass}.
For each of the distances and densities, we have plotted the lowest reliable mass bin.
The scatter in the plot reflects the random nature of the simulations.
The positions of the stars in each of the stellar fields were randomised, the sampling of the mass function was random, and the detector and shot noise were applied to the image as part of the SimCADO read-out process.
Thus no two stellar fields were the same.
Each stellar field configuration was only run once.
We therefore only have one data point for each density and distance.
The bin size used for the reliability statistics was set to 0.2 dex and is the uncertainty in the limiting observable mass.

Figure~\ref{fig:trusted_mass} immediately shows the two limiting regimes of sensitivity and crowding.
The flat parts of the curves in Figure~\ref{fig:trusted_mass} show the densities for which MICADO will be sensitivity limited at each distance, and the diagonal regions show where crowding becomes the limiting factor.
For example, observations of a cluster at a distance of 8\, kpc will always be crowding limited for densities above 100\,\spae.
At a distance of 200\,kpc, observations will be limited by sensitivity up to a density of 10\h4\,\spa, thereafter, crowding will be the dominant factor.
At 5\, Mpc, all observations will be sensitivity limited.
As a reference, we included the approximate stellar densities for three well-known young clusters in Figure~\ref{fig:trusted_mass} \textit{\textup{assuming they were located at the distance of the simulated clusters}}.
For example, if the YMC Westerlund 1 were to be located in the LMC, it would fall into the crowding-limited regime for MICADO\@.
The lowest reliably observable mass in the densest region of the core would only be \s0.5\,\msune.
This is equivalent to what HST is capable of observing in the outer rim territories of LMC clusters.
For clusters in the LMC with stellar densities lower than 10\h3~\spa , MICADO will be limited by sensitivity to masses above 0.1\,\msune.
While this mass is only 0.3\,\msun lower than what current Hubble observations can achieve, it should be emphasised that this increase of ``only'' 0.3\,\msun will reveal the majority of M-type stars, which account for almost three-quarters of all main-sequence stars~\citep{ledrew2001}.
The limit of current studies is close to the 0.5\,\msun knee from \citet{kroupa2001}, which means that opening up this range will allow future studies to precisely determine the shape of the IMF in the dense cores of young LMC clusters.

As previously noted, the exposure time for the simulated images was one hour.
By observing for longer times, the lowest observable mass will decrease.
However, the change is disproportionate to the exposure time.
Additional ``observations'' with SimCADO showed that increasing the exposure time from 1 to 10 hours per cluster only increases the sensitivity limit by around 1.5\m and 1\m in the J and K$_S$ filters, respectively.
For the case of the LMC, this would decrease the lowest observable mass to about 0.06\msune, which is just below the hydrogen-burning limit.

The cores of the majority of young clusters are less dense than that of Westerlund 1, and therefore the limiting observable mass will also be lower than the 0.5\,\msun mass quoted for a Westerlund 1-like YMC in the LMC\@.
With the resolving power of MICADO it will be possible to determine the extent to which apparent mass segregation has played a role in previous studies of the IMF in the LMC~\citep{Ascenso2009-de}.
More to the point, MICADO will enable us to understand the apparent deviations from the Salpeter IMF as reported by \citet{dario2009}, \citet{geha2013}, and \citet{kalirai2013}.

At distances of 100\,kpc to 200\,kpc and with careful photometry and longer observations, MICADO is expected to be able to detect stars down to the sensitivity limit of 0.5\,\msun.
This will only be possible for stellar densities lower than 10\h4 \spae, however.
As a reference, an ONC-like cluster at a distance of 200\,kpc would have a stellar density of about 10\h5 \spae.
Such observations would be useful for determining the composition of OB associations and sparser (older) open clusters, if there were any present in the non-Magellanic satellites of the Milky Way.
Nevertheless, MICADO will still allow us to observe the power-law break at 0.5\,\msun in the field population of the nearest low-metallicity dwarf spheroidal galaxies.

Closer to home, MICADO is expected to be able to detect 10\,M$_{Jup}$ objects in ONC-like clusters at a distance of 8\,kpc (along low-extinction lines of sight).
An obvious candidate for studies of the IMF in extreme environments is the Arches cluster because of its proximity to the Galactic centre.
The main hindrance to deep observations of the Arches cluster is, somewhat counterintuitively, not the \textgreater2 magnitudes of variable Ks-band extinction along the line of sight~\citep{espinoza2009}, but rather the \textgreater350 stars in the cluster~\citep{galacticnucleaus} that are brighter than the saturation limit for MICADO\footnote{A MICADO internal analysis shows that point sources with magnitudes K$_S$\textgreater14.8\,\m will saturate the MICADO detectors within the 2.6\,s minimum exposure time.}.
Very few regions in the cores of Milky Way open clusters do not contain stars brighter than the saturation limit.
This will make deep MICADO observations of these regions difficult.

\subsection{Core densities of young star clusters}
  \label{subsec:core_densities}

The instrumental effect that plays the most significant role by far for the accuracy of the estimates given here is our knowledge of the PSF\@.
For this study, we used a single SCAO PSF\@.
We assumed that the PSF orientation stayed the same for the length of the observation.
Consequently, we had a very good model of our reference star for the PSF subtraction.
This will not be the case for real observations as the pupil of the telescope will rotate with respect to the sky, causing an axial broadening of the PSF over the course of an observing run.
This broadening may improve the results from our subtraction method as it will smooth out many of the sharp features of the instantaneous PSF that lead to false-positive detections.
Information on both the structure of the PSF and the extent of the wings will be lost because of the rotational broadening, however.
Thus the PSF subtraction algorithm will be able to estimate the background level less accurately when the reference PSF is fit to a star.
As a consequence, faint stars caught in the PSF wings of the brighter stars may not be detected as often as they would be if the PSF remained rotationally aligned with the sky.
To extract as many stars as possible given the shape of the ELT PSF, we propose the following hybrid approach: Subtract the brightest stars from each exposure using an instantaneous PSF derived from the brightest stars in that exposure, then stack the residual images and extract the faintest stars using a rotationally broadened PSF\@.
Further investigation is required to determine whether this approach would indeed increase the detection rate for faint stars.

Regardless of the PSF shape, it is clear from our simulations that resolving stellar densities of 10\h3\,\spa is well within the capabilities of MICADO\@.
With an optimised PSF fitting and subtraction algorithm, extracting upward of 5$\times$10\h3\,\spa is also expected to be in the realm of possibility.
This is equivalent to approximately one star in an area equivalent to \s2.5 ELT H-band PSF FWHMs.
This is similar to being able to resolve every star in the core of an ONC-like cluster in the LMC\@.
For the JWST and HST, the equivalent stellar densities are only 160~\spa and 20~\spae, respectively.
Although MICADO may not have the sensitivity of a space-based telescope, the resolving power will give us full access to the core populations of dense stellar clusters in the major satellites of the Milky Way.

\subsection{Opportunities and targets for future observations with MICADO and the ELT}
  \label{subsec:future_opportunities}

These simulations are a helpful theoretical exercise.
However, without an application to observations, they are not all that useful.
Figure~\ref{fig:star_density_vs_age} shows the estimated stellar densities in the cores of the open clusters and YMCs compiled by \citet{portegies2010}.
The density values, log$_{10}$($\rho$), only take the stars with apparent magnitudes above the sensitivity limit of MICADO into account and thus reflect the ``real'' observable density for the clusters (also listed in Table~\ref{tbl:pz10_selection}).
The limits set for the HST, JWST, and MICADO are the critical stellar density above which our extraction algorithm struggles to detect and remove more than 90\% of the stars in a field.
We find that for the Galactic clusters, the resolution of the JWST will be sufficient to resolve all stars in most cluster cores down to the sensitivity limit of the instrument.

For clusters in the Galactic plane, JWST observations will struggle to distinguish the cluster stars from the field stars.
To reliably determine cluster membership, observations of the proper motion of the cluster relative to the field are required.
\citet{stolte2008} showed that the proper motion of the Arches cluster near the Galactic centre is \s5\,mas yr\h{-1}.
This equates to about one-sixth the size of a pixel in the JWST NIRCam instrument.
MICADO, in contrast, will have a plate scale of 1.5\,mas in the high-resolution mode, meaning that cluster membership could be determined by observations spaced only several months apart.
Greater certainty regarding cluster membership will significantly increase the accuracy of estimates of the cluster IMF based on star counts.

Farther away, resolving the cores of the massive young clusters in the Magellanic Clouds will not be possible with the JWST. MICADO, however, will enable access to these cluster cores, which in turn will open up the possibility of studying the dynamical processes (e.g., evaporation and core-collapse) involved in the evolution of extragalactic clusters.
Additionally, observations of a series of LMC clusters with varying ages will give a much better picture of the evolution of the initial mass function into the present-day mass function, and of the effect of the dynamical evolution of the cluster on observations and calculations of a cluster IMF\@.

\citet{portegies2010} provided a curated list of the most well-known massive young clusters inside and outside the Milky Way.
However, many more clusters exist within the Local Group of galaxies.
To fully understand the environmental dependence on cluster formation and evolution, a statistically significant number of extragalactic clusters will need to be observed.
Figure~\ref{fig:local_group_cluster_number} shows the pool of clusters that is available to MICADO at the ELT along with the corresponding observational limits of stars of various masses.
Within the Milky Way alone, over 500 star clusters become available for which a fully resolved IMF (including the brown dwarf regime) might be determined.
If the Magellanic Clouds are included for studying the region on either side of the IMF peak, this number increases to over 1000 young cluster targets.
Out to 2Mpc, the resolved high-mass slope of the IMF might be studied in between 1500 and 2500 clusters, including up to 1000 new previously undocumented star clusters\footnote{This estimate is based on the total H$_\alpha$ flux of each galaxy, with the conversion to an approximate number of clusters based on the H$_\alpha$ derived star formation rate and young cluster catalogue for M31~\citep{caldwell09}.}.

MICADO will enable IMF studies of fully resolved populations to move from direct IMF measurements of single clusters in and around the solar neighbourhood to statistically significant numbers of clusters in a diverse set of environments both inside and outside the Milky Way.
Such a statistically meaningful sample of resolved IMFs will hopefully enable a reliable determination of any environmental parameters that affect the star formation process, thus answering the major open question of IMF universality.

%%%%%%%%%%%%%%%%%%%%%%%%%%%%%%%%%%%%%%%%%%%%%%%%%%%%%%%%%%%%%%%%%%%%%%%%%%%%%%%%
% CONCLUSIONS
%%%%%%%%%%%%%%%%%%%%%%%%%%%%%%%%%%%%%%%%%%%%%%%%%%%%%%%%%%%%%%%%%%%%%%%%%%%%%%%%

\section{Conclusion}
\label{sec:conclusion}

MICADO and the ELT will provide the opportunity to resolve the core populations of dense star clusters in the Local Group.
With MICADO observations of young stellar clusters, we will finally be able to answer the question whether the IMF distribution is indeed universal, or if its shape changes outside the solar neighbourhood.
Currently, these answers are locked inside the dense cores of young stellar clusters because observations of these clusters are primarily limited by confusion.

This work reports the results of a preliminary investigation into the mass limits for future studies of the IMF in young stellar clusters in the Milky Way and the Local Group of galaxies.
To find these limits, we used the instrument simulator for MICADO (SimCADO) to generate synthetic observations of 56 dense stellar regions corresponding to the cores of young stellar clusters at varying distances from Earth.
Here we present a summary of the results.

\begin{enumerate}
    \item We have shown that MICADO will easily be able to resolve all members of a stellar population with a density up to 10\h3\,\spae.
    With proper knowledge of the PSF and an optimised detection and subtraction algorithm, densities of 5$\times$10\h3\,\spa are also expected to be achievable.

    \item Observations with MICADO will enable direct (resolved) observations of the IMF in well over 1500 young clusters in diverse environments within the Local Group of galaxies.
    These observations will provide a statistically meaningful sample for reliably quantifying possible IMF variations and the role of the environment on the shape of the IMF distribution function.

    \item The resolution of MICADO will allow the peak of the IMF (0.1\,\msun\textless M \textless0.5\,\msune) to be extensively investigated in the Magellanic Clouds, and the Salpeter slope of the high-mass region of the IMF to be studied out to distances of 5\,Mpc (M83).

    \item Observations focusing on the IMF of clusters in the LMC will be limited by sensitivity, not crowding, to 0.1\,\msune.

    \item Investigations of the transition around the hydrogen-burning limit (\s0.08\,\msune) will not be possible outside the Milky Way.
    Instead, brown dwarf populations will be accessible in the cores of the densest Milky Way clusters, for instance,\ in the Westerlund clusters or the emerging W49A clusters.
    Objects with masses as low as 10\,M$_{Jup}$ will be accessible by MICADO for clusters within 8\,kpc of Earth.
    The only caveat is that an appropriate observation strategy must be found to mask the many bright (m$_{Ks}<15^m$) stars present in all Milky Way clusters.

    \item Finally, accurate knowledge of the ELT PSF will be absolutely essential for good photometry and PSF subtraction algorithms.
    The sharp structures created by the segmented mirror design will lead to many false low-luminosity star detections if either the PSF is not well known or the extraction algorithm is not capable of differentiating between a star and an artifact of the PSF\@.

\end{enumerate}

%%%%%%%%%%%%%%%%%%%%%%%%%%%%%%%%%%%%%%%%%%%%%%%%%%%%%%%%%%%%%%%%%%%%%%%%%%%%%%%%
% Acknowledgements
%%%%%%%%%%%%%%%%%%%%%%%%%%%%%%%%%%%%%%%%%%%%%%%%%%%%%%%%%%%%%%%%%%%%%%%%%%%%%%%%

\begin{acknowledgements}

KL would also like to express his gratitude to Gijs Verdoes Kleijn, Eline Tolstoy, Ric Davies, and Joana Ascenso for the insightful and helpful comments and discussions regarding future possible observations with the ELT\@.
SimCADO incorporates Bernhard Rauscher's HxRG Noise Generator package for python~\citep{nghxrg}.
This research made use of Astropy, a community-developed core Python package for astronomy~\citep{astropy, astropy2}.
This research made use of POPPY, an open-source optical propagation Python package originally developed for the James Webb Space Telescope project~\citep{poppy}.
This research made use of Photutils~\citep{photutils}.
This research has made use of ``Aladin sky atlas'' developed at CDS, Strasbourg Observatory, France~\citep{aladin, aladinlite}.
SimCADO makes use of atmospheric transmission and emission curves generated by ESO's SkyCalc service, which was developed at the University of Innsbruck as part of an Austrian in-kind contribution to ESO\@.

This research is funded by the project IS538003 of the Hochschulraumstrukturmittel (HRSM) provided by the Austrian Government and administered by the University of Vienna.

\end{acknowledgements}

% %%%%%%%%%%%%%%%%%%%%%%%%%%%%%%%%%%%%%%%%%%%%%%%%%%%%%%%%%%%%%%%%%%%%%%%%%%%%%%%%
% % BIBLIOGRAPHY
% %%%%%%%%%%%%%%%%%%%%%%%%%%%%%%%%%%%%%%%%%%%%%%%%%%%%%%%%%%%%%%%%%%%%%%%%%%%%%%%%

\bibliographystyle{aa}
\bibliography{38145corr}

%%%%%%%%%%%%%%%%%%%%%%%%%%%%%%%%%%%%%%%%%%%%%%%%%%%%%%%%%%%%%%%%%%%%%%%%%%%%%%%%
% APPENDIX
%%%%%%%%%%%%%%%%%%%%%%%%%%%%%%%%%%%%%%%%%%%%%%%%%%%%%%%%%%%%%%%%%%%%%%%%%%%%%%%%

\begin{appendix}

%%%%%%%%%%%%%%%%%%%%%%%%%%%%%%%%%%%%%%%%%%%%%%%%%%%%%%%%%%55
% Core size and density from Portegeis-Zwart 2010
\section{Supplementary material}
\label{sec:appendix}

\begin{table*}
    \centering
    \caption{Properties of a selection of nearby YMCs from \citet{portegies2010}}
    \label{tbl:pz10_selection}
    \begin{tabular}{l l r r r r r r}
        \hline\hline
        Galaxy & Cluster      & Distance & Age  & log(mass) & Core radius & log$_{10}$($\rho$)    & Limiting mass \\
              &              & kpc      & Myr  & \msun     & arcsec  & stars arcsec\h{-2} & \msun         \\
        \hline
        \multicolumn{8}{c}{Cores resolvable by the HST}                                                     \\
        \hline
        MW     & ONC          & 0.4      & 1    & 3.7       & 100     & -1.6           & 0.01          \\
        \hline
        \multicolumn{8}{c}{Cores resolvable by the JWST}                                                    \\
        \hline
        MW     & Trumpler-14  & 2.7      & 2    & 4         & 10.7    & 1.1            & 0.01          \\
        MW     & Quintuplet   & 8.5      & 4    & 4.0       & 24      & 1.1            & 0.04          \\
        MW     & NGC3603      & 3.6      & 2    & 4.1       & 8.6     & 1.2            & 0.01          \\
        MW     & Westerlund-1 & 5.2      & 3.5  & 4.5       & 15.9    & 1.7            & 0.01          \\
        LMC    & NGC2214      & 50       & 39.8 & 4.0       & 7.5     & 1.9            & 0.1           \\
        \hline
        \multicolumn{8}{c}{Cores resolvable by MICADO}                                                  \\
        \hline
        LMC    & NGC1847      & 50       & 26.3 & 4.4       & 7.1     & 2.2            & 0.1           \\
        LMC    & NGC2157      & 50       & 39.8 & 4.3       & 8.2     & 2.2            & 0.1           \\
        LMC    & NGC1711      & 50       & 50.1 & 4.2       & 7.9     & 2.2            & 0.1           \\
        LMC    & NGC1818      & 50       & 25.1 & 4.4       & 8.5     & 2.3            & 0.1           \\
        LMC    & NGC2164      & 50       & 50.1 & 4.2       & 6.1     & 2.3            & 0.1           \\
        SMC    & NGC330       & 63       & 25.1 & 4.6       & 7.7     & 2.4            & 0.15          \\
        LMC    & NGC2136      & 50       & 100  & 4.3       & 6.6     & 2.4            & 0.1           \\
        MW     & Arches       & 8.5      & 2    & 4.3       & 4.9     & 2.5            & 0.04          \\
        LMC    & NGC1850      & 50       & 31.6 & 4.9       & 11      & 2.5            & 0.1           \\
        LMC    & NGC2004      & 50       & 20   & 4.4       & 5.8     & 2.5            & 0.1           \\
        LMC    & NGC2100      & 50       & 15.8 & 4.4       & 4.1     & 2.7            & 0.1           \\
        M31    & B257D        & 780      & 79.4 & 4.5       & 0.8     & 3.4            & 0.9           \\
        \hline
        \multicolumn{8}{c}{Only outer regions resolvable by MICADO}                                    \\
        \hline
        LMC    & R136         & 50       & 3    & 4.8       & 0.41    & 3.7            & 0.1           \\
        M31    & B066         & 780      & 70.8 & 4.3       & 0.10    & 4.2            & 0.9           \\
        M31    & B040         & 780      & 79.4 & 4.5       & 0.15    & 4.3            & 0.9           \\
        M31    & B043         & 780      & 79.4 & 4.4       & 0.19    & 4.3            & 0.9           \\
        M31    & B318         & 780      & 70.8 & 4.4       & 0.05    & 4.4            & 0.9           \\
        M31    & B448         & 780      & 79.4 & 4.4       & 0.05    & 4.4            & 0.9           \\
        M31    & Vdb0         & 780      & 25.1 & 4.9       & 0.37    & 4.4            & 0.9           \\
        M31    & B327         & 780      & 50.1 & 4.4       & 0.05    & 4.5            & 0.9           \\
        M31    & B015D        & 780      & 70.8 & 4.8       & 0.06    & 4.6            & 0.9           \\
        \hline
    \end{tabular}
    \tablefoot{The age and observable stellar densities for a selection of YMCs found both inside and outside the Milky Way, as listed in \citet{portegies2010}.
    Only clusters from \citet{portegies2010} with a defined core radius, r$_c$, are shown.
    The densities were calculated to include only stars brighter than Ks=28$^m$ because fainter stars will not be detectable by MICADO. The table lists the parameters for the clusters shown in Fig.~\ref{fig:star_density_vs_age}.}
\end{table*}

%%%%%%%%%%%%%%%%%%%%%%%%%%%%%%%%%%%%%%%%%%%%%%%%%%%%%%%%%%%%%%%%%%%%%%%%%%%%%%%%
% References for the cumulative cluster numbers within 2 Mpc

\begin{table*}
    \centering
    \caption{Number of star-forming clusters in nearby galaxies}
    \label{tbl:cum_cluster_refs}
    \begin{tabular}{lrrrc}

    \hline
    \hline
    Name     & SFR                  & Distance  & N Clusters & Reference      \\
             & [M$_\odot$ / yr]     & [kpc]     &            &                \\
    \hline
    MilkyWay &                      & 8         & 590        & 1            \\
    LMC      & 0.30                 & 50        & 456        & 2            \\
    SMC      & 0.05                 & 63        & 71         & 2            \\
    NGC6822  & 0.01                 & 499       & 24         & 3            \\
    M33      & 0.36                 & 847       & 208        & 4            \\
    NGC0055  & 0.45                 & 2128      & 168        & 5            \\
    NGC0300  & 0.18                 & 2148      & 117        & 6             \\
    NGC4214  & 0.15                 & 2938      & 52         & 7             \\
    \hline
    \end{tabular}
    \tablefoot{References for the cumulative cluster numbers within a 2 Mpc radius that will be observable by MICADO at the ELT.
    Star formation rate estimates are based on the integrated galaxy H$_{\alpha}$ fluxes from \citet{karachentsev2013} and were used to estimate the true number of open clusters contained in each galaxy (see main text).
    The Milky Way clusters were taken from the HEASARC Milky Way Star Cluster catalogue~\citep{heasarc_mwsc} and filtered to include only clusters visible to MICADO at the ELT (i.e.,\ -85~\textless~Dec~\textless~+35 deg).}
    \tablebib{(1)~\citet{heasarc_mwsc}; (2) \citet{glatt2010}; (3) \citet{karampelas2009};
              (4) \citet{fan2014}; (5) \citet{castro2008}; (6) \citet{pietrzynski2001};
              (7) \citet{andrews2013}.}

\end{table*}

\begin{figure*}
    \centering
    \includegraphics[width=\textwidth]{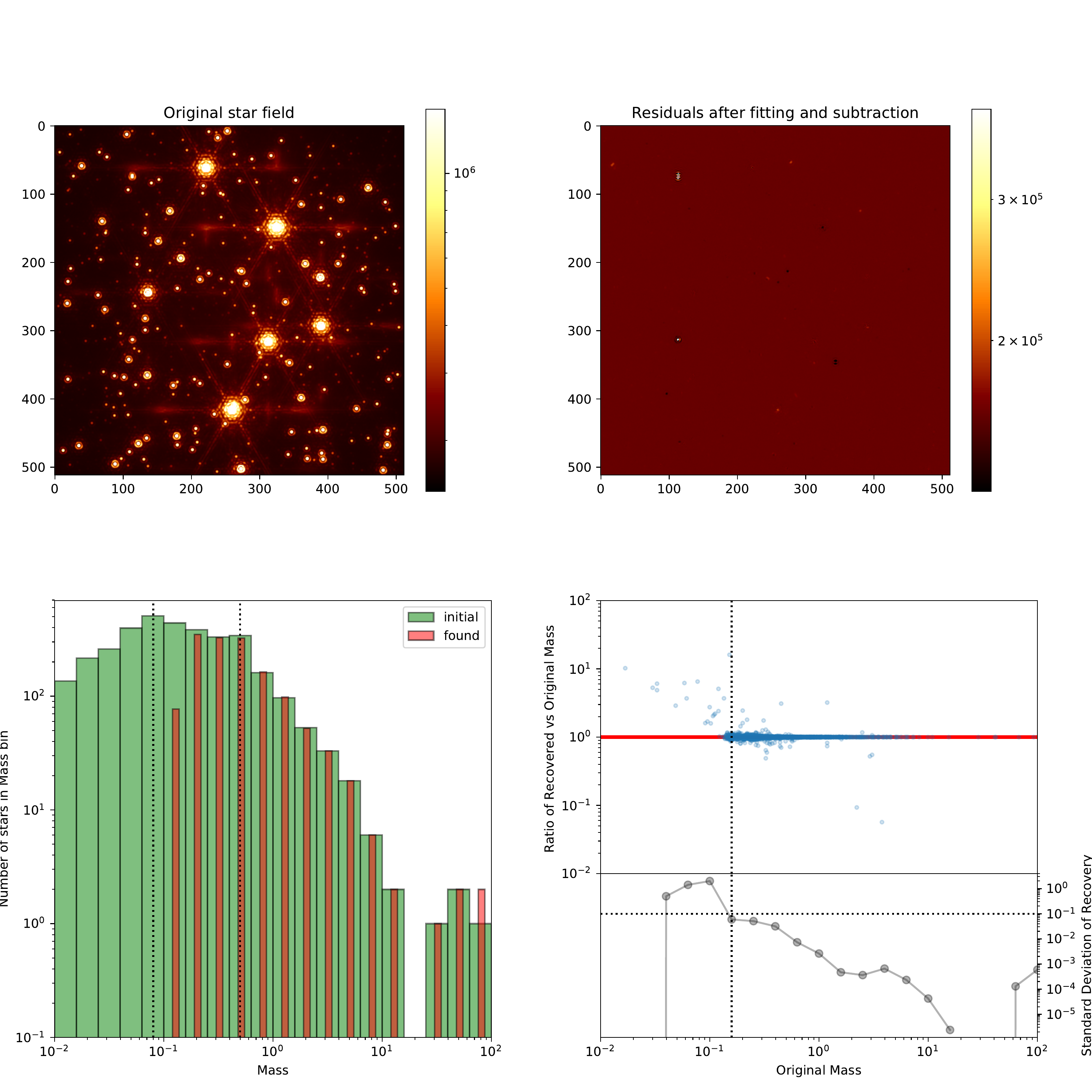}
    \caption{Results of extracting stars from a 1000\,\spa cluster at a distance of 50\,kpc.
    Top left: Original 2''$\times$2'' stellar field with a density of 10\h3~\spa.
    The stars in the field have masses between 0.01\,\msun and 300\,\msune.
    The PSF used in this study was an instantaneous SCAO PSF, similar to what would be seen on a single MICADO detector 2.6\,s exposure.
    Top right: Same field after our detection and subtraction algorithm iteratively removed all the stars.
    10\h3~\spa are extracted reasonably easily by our algorithm.
    Bottom left: Fraction of extracted stars in each mass bin that matched the original list of stars.
    The majority of stars more massive than 0.1\,\msun were detected.
    Bottom right: Upper panel: Ratio of extracted mass to original mass.
    The vast majority (\s97\%) of the almost 4000 stars in the image fell almost perfectly on the red one-to-one line.
    The minor scatter around the line arises because our detection algorithm is unable to distinguish between two very close stars and contamination from the PSF artifacts, e.g.,\ the segmented diffraction spikes.
    The lower panel shows the standard deviation of masses around the one-to-one line in a certain mass bin.
    A mass bin was deemed reliable when the average ratio of recovered to original mass was in the range 1$\pm$0.1 and the standard deviation was lower than 10\%.}

    \label{fig:results_lmc_1E3}
\end{figure*}

\begin{figure*}

    \centering
    \includegraphics[width=\textwidth]{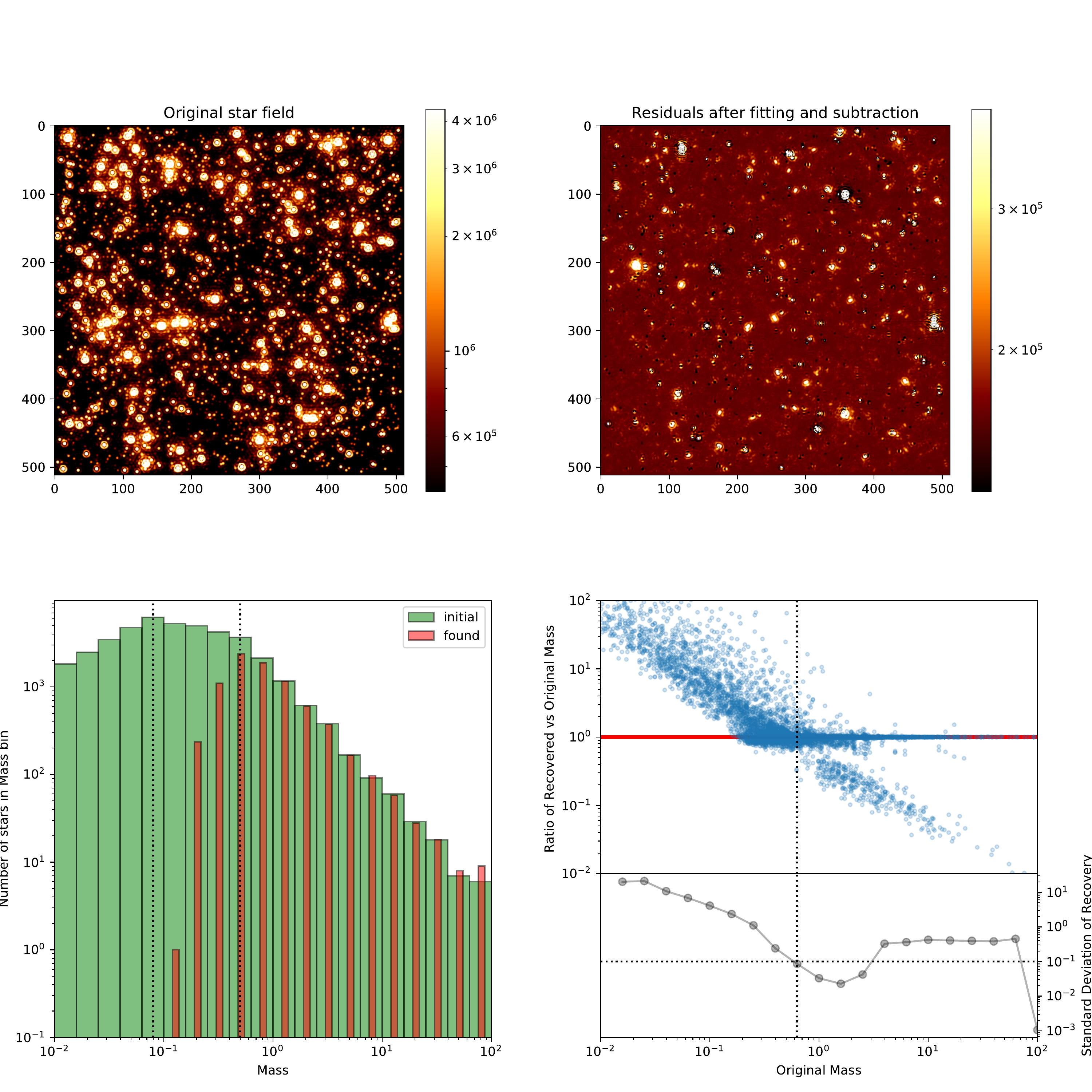}
    \caption{Same as Fig.~\ref{fig:results_lmc_1E3}, but for a stellar density of 10\h4~\spae.
    At these densities, the number of double stars has increased to the point where our detection algorithm was unable to accurately fit and subtract many of the bright stars.
    Although a large number of incorrect mass determinations are visible in the large blue cloud, about 60\% of the \s40\,000 sources in this image still fall on the red one-to-one line.
    The segmented PSF meant that the algorithm detected many false sources, which skewed the detection statistics in both the high- and low-mass regimes.
    We are still investigating ways of preventing this from happening in future studies.}

    \label{fig:results_lmc_1E4}

\end{figure*}

\end{appendix}

%________________________________________________________________

\end{document}